\documentstyle[prb,eqsecnum,aps]{revtex}

\begin{document}

\title{Many-body approach to the nonlinear interaction of charged particles
with an interacting free electron gas}
\author{T. del R\'\i o Gaztelurrutia$^{1}$ and J. M. Pitarke$^2$}
\address{$^1$ Fisika Aplikatua Saila, Industri eta Telekomunikazio Ingeniarien
Goi Eskola Teknikoa, Urkijo Zumarkalea z/g, S-48013 Bilbo, Spain\\ 
$^2$ Materia Kondentsatuaren Fisika Saila, Zientzi Fakultatea, 
Euskal Herriko Unibertsitatea,\\ 644 Posta kutxatila, 48080 Bilbo, Basque 
Country, Spain\\
and Donostia International Physics Center (DIPC) and Centro Mixto
CSIC-UPV/EHU,\\ Donostia, Basque Country, Spain}

\date{\today}

\maketitle

\begin{abstract}
We report various many-body theoretical approaches to the nonlinear
decay rate and 
energy loss of charged particles moving in an interacting
free electron gas. These include perturbative formulations of the scattering
matrix, the self-energy, and the induced electron density. Explicit expressions
for these quantities are obtained, with inclusion of exchange and
correlation effects. 
\end{abstract}


\section{Introduction}

The energy loss of non-relativistic charged particles entering a
metal is primarily due to the creation of electron-hole pairs and collective
excitations in the solid, interactions with the nuclei only becoming important
when the velocity of the projectile is much smaller than the mean speed of the
electrons in the solid\cite{Echenique}.
 
The degenerate interacting free electron gas (FEG) provides a good
model to describe a regime in which electrons are responsible for the
energy-loss process. The inelastic decay rate and energy loss
of charged particles in a FEG have been calculated for many years in the
first-Born approximation or, equivalently, within linear response theory. It is
well known that these first-order calculations predict an energy-loss that grows
with the square of the projectile charge, $Z_1e$, and provide a good
approximation when the velocity of the projectile is much larger that the
average velocity of the target electrons. However, when the velocity of the
projectile decreases non-linearities become apparent. An important
example is provided by the existing differences between the energy loss of
protons and antiprotons\cite{antiproton,Moller0,Moller}, which cannot be
accounted for within linear-response theory. These differences were then
successfully accounted on the basis of second-order perturbative calculations
that used the random-phase approximation (RPA) and treated the moving charged
particle as a prescribed source of energy and
momentum.\cite{Sung,Zaremba,Esbensen,Pitarke1,Pitarke2} Beyond-RPA
calculations of this so-called $Z_1^3$ effect have been reported only very
recently, in the limit of low velocities.\cite{RP00} 

In this paper, we report various many-body theoretical approaches to the
quadratic decay rate and energy loss of charged particles moving in an
interacting FEG, which include exchange-correlation (xc) effects and
treat the moving charged particle as part of the many-body interacting system.
First of all, we present a fully quantum treatment of the probe particle, which
we assume to be distinguishable from the electrons in the Fermi gas. We assign a
propagator to this particle, and then follow procedures of many-body
perturbation theory to derive explicit expressions for the scattering matrix.
From the knowledge of this matrix, both the decay rate and the energy loss of
the moving particle can be evaluated either within the RPA or by including
short-range xc effects. We also derive an explicit 
expression for the self-energy of the probe particle, which enables us
to present an alternative derivation of the
decay rate. 

The rest of this paper is organized as follows: A diagrammatic analysis of the
decay rate of a moving charged particle in a FEG is presented in
Section II. The decay rate and stopping power are calculated up to third order in the projectile
charge from the knowledge of the scattering matrix. It is shown that
for a heavy projectile
the decay rate agrees with the imaginary part of the projectile
self-energy, and that the the stopping power agrees with the result of
quadratic-response theory. Our
conclusions are presented in Section III. 
Atomic units
are used throughout, i.e., $e^2=\hbar=m_e=1$.   

\section{Diagrammatic analysis}

We consider the interaction of a moving probe particle of charge $Z_1$ and mass
$M$ with a FEG of density $n$. The probe particle is assumed to be
distinguishable from the electrons in the Fermi gas, which is described by
an isotropic homogeneous assembly of interacting electrons immersed in a uniform
background of positive charge and volume $V$.

In the representation of second quantization, the interaction-picture 
perturbing Hamiltonian
reads
\begin{eqnarray}\label{intham}
H_I'(t)=&&-Z_1\int d^3{\bf r}\,d^4X
\,\psi^{\dagger}(x)\,\psi(x)\,v(x,X)\,\tilde\psi^{\dagger}(X)\,\tilde\psi(X)
\cr\cr
&&+\,{1\over2}\int d^3{\bf r}\,d^4x'
\,\psi^{\dagger}(x)\,\psi(x)\,v(x,x')\,\psi^{\dagger}(x')\,\psi(x')+H_I^{{\text
BG}},
\end{eqnarray}
where $v(x,x')$ [$x=({\bf r},t)$] is the instantaneous Coulomb interaction and
the last term represents the interaction of electrons and probe
particle with the positive background. The field operators $\psi(x)$
and $\psi^{\dagger}(x)$ destroy
and create an electron at time
$t$ and point ${\bf r}$, while $\tilde\psi(X)$ and
$\tilde\psi^{\dagger}(X)$ destroy and create
the probe particle at time $t$ and point ${\bf R}$. Annihilation operators can be
written as
\begin{equation}\label{field1}
\psi(x)=\sum_i{\rm e}^{i\,\omega_i\,t}\,\phi_i({\bf r})\,a_i
\end{equation}
and 
\begin{equation}\label{field2}
\tilde\psi(X)=\sum_i{\rm e}^{i\,\omega_i\,t}\,\tilde\phi_i({\bf R})A_i,
\end{equation}
where  the operators $a_i$ and $A_i$ annihilate an electron and the
probe particle in the one-particle free states $\phi_i({\bf r})$ and
$\tilde\phi_i({\bf R})$ of energy $\omega_i$. As we are dealing
with a homogeneous system, these states can be taken to be plane-wave
states. We choose states of
momentum ${\bf k}$ and energy $\omega_{\bf k}={\bf k}^2/2$ for
electrons, and momentum ${\bf p}$ and energy
$\omega_{\bf p}={\bf p}^2/(2M)$ for the probe particle. 

The scattering matrix can be written as a time-ordered
exponential\cite{Fetter}
\begin{equation}\label{smatrix}
S=T\left\{\exp\left[-i\int^{\infty}_{-\infty}dt
\,{\rm e}^{-\eta\,|t|}\,H'_{I}(t)\right]\right\},
\end{equation}
where  $H'_I$ is the perturbing Hamiltonian of Eq. (\ref{intham}), $T$ is the
chronological operator, and $\eta$ is a positive infinitesimal.

In the interaction picture, electron and probe-particle propagators
can be expressed as
\begin{equation}\label{2pointG}
G(x,x')=-i\,
{\left\langle 0,\Phi_0\right|T\,\psi_I(x)\,\psi^{\dagger}_I(x')\,S\left|0,\Phi_0
\right\rangle\over\left\langle 0,\Phi_0\right|S\left|0,\Phi_0\right\rangle}
\end{equation}
and
\begin{equation}\label{2pointD}
D(X,X')=-i\,
{\left\langle 0,\Phi_0\right|T\,\psi_I(X)\,\psi^{\dagger}_I(X')\,S\left|0,\Phi_0
\right\rangle\over\left\langle 0,\Phi_0\right|S\left|0,\Phi_0\right\rangle},
\end{equation}
 where $|0,\Phi_0\rangle=|0\rangle|\Phi_0\rangle$
represents the noninteracting free Fermi sea with no probe particle.
Noninteracting electron and probe-particle propagators are easily found from Eqs.
(\ref{2pointG}) and (\ref{2pointD}) to be given by the following simple
expressions:
\begin{equation}\label{ionpro1}
G^0(x,x')=-i\,
\left\langle\Phi_0\right|T\,\psi(x)\,\psi^{\dagger}(x')\left|\Phi_0\right\rangle 
\end{equation}
and
\begin{equation}\label{ionpro2}
D^0(X,X')=-i\,
\left\langle 0\right|T\,\psi(X)\,\psi^{\dagger}(X')\left|0\right\rangle, 
\end{equation}
respectively.

We note that the probe-particle propagator is a retarded function, i.e., it is
different from zero only if $t>t'$. As a consequence, probe-particle
bubbles do no contribute to the
diagrammatic expansion. Therefore, the expansion of
Eq. (\ref{2pointD}) does not depend on whether the
probe particle is a fermion or a boson, and there is no probe-particle
contribution to the denominator of Eqs. (\ref{2pointG}) and
(\ref{2pointD}).

\subsection{Scattering approach}

Let us consider the process corresponding to the creation of a single
electron-hole pair, where the system is carried from an initial state
$A^{\dagger}_i\left|0,\Phi_0\right\rangle$ to a final state
$a^{\dagger}_{f_1}a_{i_1}A_{f}^{\dagger}\left|0,\Phi_0\right\rangle$. The
scattering-matrix element for this process is
\begin{equation}\label{s}
S_{f,f_1;i,i_1}={\left\langle 0,\Phi_0\right|
a_{f_1}\,a_{i_1}^{\dagger}\,A_{f}\,S
\,A^{\dagger}_{i}\left|0,\Phi_0\right\rangle\over\left\langle 0,\Phi_0\right
|S\left|0,\Phi_0\right\rangle}.
\end{equation}

Similarly, one may consider  a
double excitation, in which the system is carried from an initial state
$A^{\dagger}_i\left|0,\Phi_0\right\rangle$ to a final state
$a^{\dagger}_{f_1}a^{\dagger}_{f_2}a_{i_1}a_{i_2}A_{f}^{\dagger}
\left|0,\Phi_0\right\rangle$.
The matrix element for this process is    
\begin{equation}\label{s2}
S_{f,f_1,f_2;i,i_1,i_2}={\left\langle 0,\Phi_0\right|
a_{f_1}\,a_{f_2}\,a_{i_1}^{\dagger}\,a_{i_2}^{\dagger}\,A_{f}\,S
\,A^{\dagger}_i\left|0,\Phi_0\right\rangle\over\left\langle 0,\Phi_0\right
|S\left|0,\Phi_0\right\rangle}.
\end{equation}

After introduction of the Hamiltonian of Eq. (\ref{intham}) into Eq.
(\ref{smatrix}), the matrix elements $S_{f,f_1;i,i_1}$ and $S_{f,f_1,f_2;i,i_1,i_2}$  of
Eqs. (\ref{s}) and (\ref{s2}) can be expanded in powers of the coupling
constant $e^2$. Then, the use of Wick's theorem yields explicit
expressions for the various contributions to this expansion, in terms of the
noninteracting propagators $G^0(x,x')$ and $D^0(X,X')$. Introducing standard
Fourier representations and taking the free-particle states to be momentum
eigenfunctions, all contributions to the scattering-matrix can be derived from
scattering-like Feynman diagrams, as follows:

\noindent 1. Draw all distinct scattering diagrams, in momentum space. All
particle lines must be directed. Different ways of directing them that are not
topologically equivalent give distinct contributions. Exclude probe-particle
bubbles.

\noindent 2. Assign momentum and energy to all particle and interaction lines,
so that the sum of the four-momenta entering a vertex equals the sum of
four-momenta leaving the vertex.

\noindent 3. Include an overall factor $2\pi\,V\,
\delta_{\bf k}\,\delta(k^0)$, which represents total momentum and energy
conservation. $\delta_{\bf k}$ is the Kronecker $\delta$ symbol and
$\delta(k^0)$ is the Dirac $\delta$ function. 

\noindent 4. For every external particle line include a factor $V^{-1/2}$.

\noindent 5. For every internal electron line include 
$i\,G^0_k$, where $G^0_k$ is the noninteracting one-electron propagator in
momentum space:
\begin{equation}
G_k^0={1-n_{\bf k}\over k^0-\omega_{\bf k}+i\,\eta}+{n_{\bf k}\over
k^0-\omega_{\bf k}-i\,\eta},
\end{equation}
$({\bf k},k^0)$ being the four-momentum of the particle, $n_{\bf k}$ 
the occupation number, [$n_{\bf k}=\Theta(q_F-|{\bf q}|)$, where $q_F$ is the
Fermi momentum], and $\omega_{\bf k}={\bf k}^2/2$.

\noindent 6. For every internal probe-particle line include a factor
$i\,D^0_p$, where $D^0_p$ is the noninteracting probe-particle propagator in
momentum space:
\begin{equation}\label{d0}
D_p^0={1\over p^0-\omega_{\bf p}+i\,\eta},
\end{equation}
$({\bf p},p^0)$ being the four-momentum of the particle, and $\omega_{\bf
p}={\bf p}^2/(2M)$.

\noindent 7. For every probe$\_$particle-electron and electron-electron
interaction line include a factor $i\,Z_1\,v_{\bf q}$ and
$-i\,v_{\bf q}$ respectively, $v_{\bf q}$ being the Fourier transform
of the bare Coulomb potential.

\noindent 8. For every electron loop include a factor $-2$.

\noindent 9. Integrate over free four-momenta, $\int d^4q/(2\pi)^4$.

Since all scattering-matrix elements include delta functions accounting for
momentum and energy conservation, one may factorize them as follows
\begin{equation}\label{eq1}
S_{f,f_1;i,i_1}=2\pi\,\delta_{{\bf p}_f-{\bf p}_i-{\bf k}_{f_1}+{\bf k}_{i_1}}\,
\delta(\omega_{{\bf p}_f}\!-\omega_{{\bf p}_i}\!-\omega_{{\bf k}_{f_1}}\!+\omega_{{\bf
k}_{i_1}})\,T_{f,f_1;i,i_1}
\end{equation}
and
\begin{equation}\label{eq2}
S_{f,f_1,f_2;i,i_1,i_2}=2\pi\,
\delta_{{\bf p}_f-{\bf p}_i-{\bf k}_{f_1}-{\bf k}_{f_2} +{\bf k}_{i_1}+{\bf
k}_{i_2}}\,\delta(\omega_{{\bf p}_f}\!-\omega_{{\bf p}_i}
\!-\omega_{{\bf k}_{f_1}}\!-\omega_{{\bf k}_{f_2}}\!+\omega_{{\bf
k}_{i_1}}\!+\omega_{{\bf k}_{i_2}})\,
\,T_{f,f_1,f_2;i,i_1,i_2},
\end{equation}
where ${\bf k}_{i_1,i_2,f_1,f_2}$ and
${\bf p}_{i,f}$ represent the initial and
final momenta of target electrons and probe particle,
respectively, with energies $\omega_{{\bf
k}_{i_1,i_2,f_1,f_2}}={\bf k}_{i_1,i_2,f_1,f_2}^2/2$ and
$\omega_{{\bf p}_{i,f}}={\bf p}_{i,f}^2/(2M)$. 

The probabilities $\gamma_q^{\rm single}$ and $\gamma_q^{\rm double}$
for the probe particle to transfer four-momentum $q$ ($q^0>0$) to a FEG by
creating single and double excitations are derived by summing the
matrix elements over all available initial and final electron states and
all final probe-particle states \cite{Pitarke1}. 
\begin{equation}\label{single}
\gamma_q^{\rm single}=4\pi\,\sum_{\bf k}\,n_{{\bf k}}\,(1-n_{\bf k+q})\, 
|T_{{\bf q},{\bf k}}(p_i)|^2\,\delta\left(q^0+\omega_{\bf k}-\omega_{\bf
k+q}\right)\,\delta\left[q^0-{\bf q}\cdot{\bf v}+q^2/(2M)\right]
\end{equation}
and
\begin{eqnarray}\label{double}
\gamma_q^{\rm double}=&&8\pi\,\sum_{{\bf q}_1}\int dq_1^0
\sum_{{\bf k}_1}\,\sum_{{\bf k}_2}\,n_{{\bf k}_1}\,
(1-n_{{\bf k}_1+{\bf q}_1})\,n_{{\bf k}_2}\,(1-n_{{\bf k}_2+{\bf q}-{\bf q}_1})
\,|T_{{\bf q},{\bf q}_1,{\bf k}_1,{\bf k}_2}(p_i)|^2\cr\cr
&&\times
\delta\left(q_1^0+\omega_{{\bf k}_1}\!-\omega_{{\bf k}_1+{\bf q}_1}\right)
\delta\left(q^0-q_1^0+\omega_{{\bf k}_2}\!-
\omega_{{\bf k}_2+{\bf q}-{\bf q}_1}\right)
\delta\left[q^0-{\bf q}\cdot{\bf v}+q^2/(2M)\right],
\end{eqnarray}
where ${\bf v}$ represents the velocity of the probe particle.

In these equations recoil has not been neglected. Moreover the quantum
character of the probe particle is implicit in the $T$-matrix
elements, which include the probe-particle propagator $D^0(p)$. Therefore they
generalize the results of Ref. \onlinecite{Pitarke1} to the
case of an arbitrary distinguishable probe particle.

Hence, the total decay rate of the probe charge is given by the following
expression:
\begin{equation}\label{tau}
\tau^{-1}(p)=\sum_{\bf q}\int_0^\infty dq^0\left[\gamma_q^{\rm single}+\gamma_q^{\rm
double}+...\right].
\end{equation}

The average energy lost per unit length traveled by the probe particle, i.e.,
the so-called stopping power of the target is obtained by inserting
$q^0/v$ inside the integrand in Eq. (\ref{tau}):  
\begin{equation}\label{stopping}
-{dE\over dx}{\Big (}p{\Big)}={1\over v}\sum_{\bf q}\int_0^\infty
dq^0\,q^0\,\left[\gamma_q^{\rm single}+\gamma_q^{\rm double}+...\right].
\end{equation}

\bigskip

It is well known that the decay rate and the energy loss cannot be computed by
simply evaluating the lowest-order tree-level Feynman diagrams, because of
severe infrared divergences due to the long-range Coulomb interaction. Instead,
one needs to resum electron-loop corrections and expand the scattering matrix in
terms of the dynamically screened Coulomb interaction. 

Direct, linear, and
quadratic contributions to the screened interaction are represented in Fig. 1.
Dashed lines [$-i\,v_{\bf q}$] represent the bare Coulomb
interaction.
The full bubble and triangle, denoted $i\,\chi_q$ and $-2\,Y_{q_1,q_2}$,
 represent the sum of all possible Feynman
diagrams joining two and three points, and thus correspond to the
Fourier transform of time-ordered density correlation functions of the
interacting FEG :
\begin{equation}\label{chi01}
\chi_q=\int d^4x_1\,{\rm e}^{-i\left[{\bf q}\cdot({\bf r}_1-{\bf
r}_2)-q^0(t_1-t_2)\right]}\,\chi(x_1,x_2)
\end{equation}
and
\begin{eqnarray}\label{chi02}
Y_{q_1,q_2}=\int d^4x_1\,\int d^4x_2\,
&&{\rm e}^{-i\left[{\bf q}_1\cdot({\bf r}_1-{\bf
r}_2)-q^0_1(t_1-t_2)\right]}\,{\rm e}^{-i\left[({\bf q}_1+{\bf
q}_2)\cdot({\bf r}_2-{\bf r}_3)-(q^0_1+q^0_2)(t_2-t_3)\right]}\,Y(x_1,x_2,x_3),
\end{eqnarray}
with
\begin{equation}\label{chi1} 
\chi(x,x')= 
-i\,
\left\langle\Psi_0|T\,\tilde\rho_H(x)\,\tilde\rho_H(x')|\Psi_0\right\rangle
\end{equation}
and
\begin{equation}\label{chi2} 
Y(x,x',x'')=-{1\over 2}
\left\langle\Psi_0|T\,\tilde\rho_H(x)\,\tilde\rho_H(x')\,\tilde\rho_H(x'')|
\Psi_0\right\rangle.
\end{equation}
Here, $|\Psi_0\rangle$ represents the normalized exact many-electron ground
state, and $\tilde\rho_H(x)=\hat\rho_H(x)-n$ is the exact Heisenberg
electron-density fluctuation operator, both in the absence of probe particle.

Introducing complete sets of energy eigenstates between the Heisenberg fields of
Eqs. (\ref{chi1}) and (\ref{chi2}), one obtains spectral representations for
$\chi_q$ and $Y_{q_1,q_2}$. We find:
\begin{equation}\label{eq17}
\chi_q=V^{-1}\sum_n\left|(\rho_{\bf
q})_{n0}\right|^2\left[{1\over q^0-\omega_{n0}+{\rm
i}\eta_{q^0}}-{1\over q^0+\omega_{n0}+i\,\eta_{q^0}}\right]
\end{equation}
and
\begin{eqnarray}\label{eq2p}
Y_{q_1,q_2}=-{1\over 2}\,V^{-1}
\sum_{n,l}&&\left[\,{(\rho_{{\bf q}_1})_{0n}(\rho_{{\bf
q}_3})_{nl}(\rho_{{\bf q}_2})_{l0}\over (q^0_1-\omega_{n0}+{\rm
i}\eta_{q^0_1})(q^0_2+\omega_{l0}+i\,\eta_{q^0_2})}+{(\rho_{{\bf
q}_2})_{0n}(\rho_{{\bf q}_1})_{nl}(\rho_{{\bf q}_3})_{l0}\over
(q^0_2-\omega_{n0}+i\,\eta_{q^0_2})(q^0_3+\omega_{l0}+{\rm
i}\eta_{q^0_3})}\right.\cr\cr
&&\left.+{(\rho_{{\bf q}_3})_{0n}(\rho_{{\bf
q}_2})_{nl}(\rho_{{\bf q}_1})_{l0}\over (q^0_3-\omega_{n0}+{\rm
i}\eta_{q^0_3})(q^0_1+\omega_{l0}+i\,\eta_{q^0_1})}+(q_2\rightarrow q_3)\right],
\end{eqnarray}
where $\eta_q=\eta\,{\rm sgn}(q^0)$, $\omega_{nl}=E_n-E_l$,
$q_3=-(q_1+q_2)$ and $\left(\rho_{\bf
q}\right)_{nl}$ is the matrix element of the Fourier transform of the
electron-density operator, taken between exact many-electron states of energy 
$E_n$ and $E_l$.

In the RPA, density correlation functions are obtained
by summing
over all ring-like diagrams, as shown in Fig. 2, thereby neglecting all
self-energy, vertex, and vertex-ladder insertions [see Fig. 3]. Hence,
\begin{equation}\label{RPA1}
\chi^{\rm RPA}_q=\chi_q^0+\chi_q^0\,v_{\bf q}\,\chi^{RPA}_q
\end{equation}
and
\begin{equation}\label{RPA2}
Y_{q_1,q_2}^{RPA}=
K_{q_1}^{RPA}\,Y_{q_1,q_2}^0\,K_{-q_2}^{RPA}\,K_{-q_3}^{RPA},
\end{equation}
where $\chi_q^0$ and $Y_{q_1,q_2}^0$ are the non-interacting FEG
density correlation functions and $K_q$ is the so-called inverse dielectric
function:
\begin{equation}\label{inverse}
K^{RPA}_q=1+\chi^{RPA}_q\,v_{\bf q}.
\end{equation}

Improvements on the RPA are typically carried out by introducing an
effective e-e interaction\cite{Singwi}
\begin{equation}\label{eqHub}
\tilde v_q=v_{\bf q}\left(1-G_q\right),
\end{equation}
where $G_q$ is the so-called local-field factor, first introduced by
Hubbard,\cite{Hubbard} accounting for all self-energy, vertex, and vertex-ladder
insertions not present in the RPA. Accordingly, the density
correlation functions $\chi_q$ and $Y_{q_1,q_2}$ are found to be of the RPA
form, but with all e-e bare-Coulomb interactions
$v_{\bf q}$ replaced by $\tilde v_q$\cite{Sayasov}:
\begin{equation}\label{chitilde}
\chi_q=\chi_q^0+\chi_q^0\,\tilde v_q\,\chi_q,
\end{equation}
and
\begin{equation}\label{Ytilde}
Y_{q_1,q_2}=
\tilde K_{q_1}\,Y_{q_1,q_2}^0\,\tilde K_{-q_2}\,\tilde K_{-q_3},
\end{equation}
where the so-called test$\_$charge-electron inverse dielectric
function has been introduced:\cite{Kleinman,HL}
\begin{equation}\label{inversep}
\tilde K_q=1+\chi_q\,\tilde v_q.
\end{equation}
This inverse dielectric function screens the potential generated by a
distinguishable test charge and 'felt' by an electron, whereas the so-called
test$\_$charge-test$\_$charge inverse dielectric function $K_q$ of Eq.
(\ref{inverse}) screens the potential both generated and 'felt' by a
distinguishable test charge:
\begin{equation}\label{inversep}
K_q=1+\chi_q\, v_q.
\end{equation}  

Now we proceed to expand the matrix elements of Eqs. (\ref{s}) and (\ref{s2})
in powers of the dynamically screened Coulomb interaction.
At this point, we will only introduce self-energy and vertex insertions that can
be described with the use of a static local-field factor [$G_q\to G_{{\bf
q},0}$]. Within this approximation, all processes corresponding to the creation
of single and double excitations can be represented by diagrams of
Fig. 4. [as in the RPA there
are no contributions, up to second order in $Z_1$, from triple and higher-order
excitations\cite{Pitarke1}]. Thus, one finds:
\begin{eqnarray}\label{tsinglep}
T_{{\bf q},{\bf k}}=&&i\,Z_1\,V^{-1}\,v_{\bf
q}\,\tilde K_q+Z_1^2\,V^{-1}\int{d^4q_1\over(2\pi)^4}
\left[2\,\tilde v_q\,v_{{\bf q}_1}\,v_{{\bf q}-{\bf q}_1}\,D^0_{p-q_1}\,
Y_{q,-q_1}\right.\cr\cr
&&\left.+v_{{\bf q}_1}\,\tilde K_{q_1}\,v_{{\bf q}-{\bf q}_1}\,\tilde K_{q-q_1}\,
D^0_{p-q_1}\left(G^0_{k+q_1}+G^0_{k+q-q_1}\right)\right]
\end{eqnarray}
and
\begin{eqnarray}\label{tdoublep}
T&&_{{\bf q},{\bf q}_1,{\bf k}_1,{\bf k}_2}=i\,Z_1\,V^{-2}\left[
2\,v_{\bf q}\,\tilde v_{q_1}\,\tilde
v_{q-q_1}\,Y_{q,-q_1}\right.\cr\cr
&&\left.+v_{\bf q}\,\tilde K_q\,
v_{{\bf q}-{\bf q}_1}\,\tilde
K_{q-q_1}\left(G^0_{k_1+q}+G^0_{k_1-q+q_1}\right)+
v_{\bf q}\,\tilde K_q\,v_{{\bf q}_1}\,\tilde
K_{q_1}\left(G^0_{k_2+q}+G^0_{k_2-q_1}\right)\right]\cr\cr
&&-i\,Z_1^2\,V^{-2}\,v_{{\bf q}_1}\,\tilde K_{q_1}\,
v_{{\bf q}-{\bf q}_1}\,\tilde K_{q-q_1}\,D^0_{p-q_1}.
\end{eqnarray}

As static local-field factors are known to be real [${\rm Im}\,G_{{\bf
q},0}=0$], one easily finds:
\begin{equation}\label{g1}
{\rm Im}\,K_q=v_{\bf q}\,|\tilde K_q|^2\,{\rm Im}\,\chi_q^0
\end{equation}
and
\begin{equation}\label{g2}
{\rm Im}\,\tilde K_q=\tilde v_q\,|\tilde K_q|^2\,{\rm Im}\,\chi_q^0,
\end{equation}
where $K_q$ and $\tilde K_q$ are the inverse dielectric functions of Eqs.
(\ref{inverse}) and (\ref{inversep}), with the density correlation function
$\chi_q$ being given in both cases by Eq. (\ref{chitilde}).

Introduction of Eqs. (\ref{tsinglep}) and (\ref{tdoublep}) into Eqs. (\ref{single})
and (\ref{double}) yields the following results, valid up to third order in the
probe$\_$particle-electron screened interaction:

\begin{eqnarray}\label{single1}
\gamma_q^{\rm single}=&&-2\,Z_1^2\,V^{-1}\,v_{\bf q}\left\{{\rm Im}K_q
+4\,Z_1\int{d^4q_1\over(2\pi)^4}\,v_{{\bf q}_1}
\,v_{{\bf q}-{\bf q}_1}\left[{\rm Im}\tilde K_q\right.\right.\cr\cr
&&\left.\left.\times\,{\rm
Im}\left(D_{p-q_1}^0\,Y_{q,-q_1}^0\,\tilde
K_{q_1}\,\tilde K_{q-q_1}\right) +{\rm
Im}\left({\tilde K_q}^*\,\tilde K_{q_1}\,\tilde K_{q-q_1}\,
D_{p-q_1}^0\,I_{q,q_1}\right)\right]\right\}\cr\cr
&&\times\,\delta\left[q^0-{\bf q}\cdot{\bf v}+q^2/(2M)\right]\Theta(q^0)
\end{eqnarray}
and
\begin{eqnarray}\label{double1}
\gamma_q^{\rm double}=&&-16\,Z_1^3\,V^{-1}\,v_{\bf q}
\int{d^4q_1\over(2\pi)^4}\,v_{{\bf q}_1}\,v_{{\bf q}-{\bf q}_1}
\left\{{\rm Im}\tilde K_{q_1}\,{\rm Im}\tilde K_{q-q_1}\,{\rm Re}
\left(\tilde K_q\,D_{p-q_1}^{0^*}\,Y_{q,-q_1}^0\right)\right.\cr\cr
&&\left.+{\rm Im}\tilde K_{q-q_1}\,{\rm
Re}\left[\tilde K_q\,{\tilde K_{q_1}}^*\,
\left(D_{p-q_1}^{0^*}+ D_{p-q+q_1}^{0^*}\right)\,I_{q_1,q}\right]\right\}\,
\delta\left[q^0-{\bf q}\cdot{\bf v}+q^2/(2M)\right]\cr\cr
&&\times\Theta(q_1^0)\,\Theta(q^0-q_1^0),
\end{eqnarray}
where we have defined the function $I_{q,q_1}$ as 
\begin{equation}\label{jota}
I_{q,q_1}={1\over 2}\left[H_{q,q_1}+H_{q,q-q_1}+i\left(J_{q,q_1}
+J_{q,q-q_1}\right)\right],
\end{equation}
with
\begin{equation}\label{hache}
H_{q,q_1}=-2\pi\,V^{-1}\sum_{\bf k}\,n_{{\bf k}}\,(1-n_{\bf k+q}) 
\left[{\delta\left(q^0+\omega_{\bf k}-\omega_{{\bf k}+{\bf q}}\right)
\over q_1^0+\omega_{\bf k}-\omega_{{\bf k}+{\bf q}_1}}
-{\delta\left(q^0-\omega_{\bf k}+\omega_{{\bf k}+{\bf q}}\right)
\over q_1^0-\omega_{\bf k}+\omega_{{\bf k}+{\bf q}_1}}\right]
\end{equation}
and
\begin{eqnarray}
J_{q,q_1}=&&2\pi^2\,V^{-1}\sum_{\bf k}\,n_{{\bf k}}\,(1-n_{\bf k+q})
\delta\left(q^0+\omega_{\bf k}-\omega_{{\bf k}+{\bf q}}\right)\cr\cr
&&\times\left[(1-n_{{\bf k}+{\bf q}_1})\,
\delta\left(q_1^0+\omega_{\bf k}-\omega_{{\bf k}+{\bf q}_1}\right)
-n_{{\bf k}+{\bf q}-{\bf q}_1}\,
\delta\left(q^0-q_1^0+
\omega_{{\bf k}}-\omega_{{\bf k}+{\bf q}-{\bf q}_1}\right)\right].
\end{eqnarray}

Introduction of Eqs. (\ref{single1}) and (\ref{double1}) into Eqs. (\ref{tau})
and (\ref{stopping}) yields the total decay rate and the average energy loss of
arbitrary particles that are distinguishable from the electrons in the Fermi
gas, with inclusion of static many-body local-field effects.

\bigskip

In order to compare our result with previous work, we consider now the
case where the probe particle is very heavy ($M>>1$) and recoil can be neglected, i.e.,
$\omega_{\bf p}-\omega_{{\bf p}-{\bf q}}\sim{\bf q}\cdot{\bf v}$.
 It this approximation, the principal
part of the noninteracting probe-particle propagator is found to give no
contribution to the integrals of Eqs. (\ref{single1}) and (\ref{double1}), and
one finds 
\begin{eqnarray}\label{single2}
\gamma_q^{\rm single}&&=2\,Z_1^2\,V^{-1}\,v_{\bf
q}\left\{-{\rm Im}K_q+4\pi\,Z_1\int{d^4q_1\over(2\pi)^4}\,v_{{\bf q}_1}
\,v_{{\bf q}-{\bf q}_1}\,\delta\left(q_1^0-{\bf q}_1\cdot{\bf
v}\right)\right.\cr\cr
&&\left.\times\left[{\rm Im}\tilde K_q\,{\rm Re}\left(Y_{q,-q_1}^0\,\tilde
K_{q_1}\,\tilde K_{q-q_1}\right)
+{\rm Re}\left({\tilde K_q}^*\,\tilde K_{q_1}\,\tilde
K_{q-q_1}\,I_{q,q_1}\right)\right]
\right\}\,\delta\left(q^0-{\bf q}\cdot{\bf v}\right)\Theta(q^0)
\end{eqnarray}
and
\begin{eqnarray}\label{double2}
\gamma_q^{\rm double}&&=16\pi\,Z_1^3\,V^{-1}\,v_{\bf q}
\int{d^4q_1\over(2\pi)^4}\,v_{{\bf q}_1}\,v_{{\bf q}-{\bf q}_1}
\left[{\rm Im}\tilde K_{q_1}\,{\rm Im}\tilde K_{q-q_1}\,{\rm Im}
\left(\tilde K_q\,Y_{q,-q_1}^0\right)\right.\cr\cr
&&\left.+2\,{\rm Im}\tilde K_{q-q_1}\,{\rm
Im}\left(\tilde K_q\,{\tilde K_{q_1}}^*\,I_{q_1,q}\right)\right]\,
\delta\left(q_1^0-{\bf q}_1\cdot{\bf v}\right)\,
\delta\left(q^0-{\bf q}\cdot{\bf v}\right)\,\Theta(q_1^0)\,\Theta(q^0-q_1^0).
\end{eqnarray}
These decay probabilities, which account for the
existence of many-body static local-field corrections, coincide in the RPA
[$G_q=0$] with those derived in Ref.\onlinecite{Pitarke2}
by treating the probe particle as a prescribed source of energy and momentum.

Simplified expressions for the total decay rate and the average energy loss of
heavy ($M>>1$) probe particles can be obtained with the aid of the following
relationship, obtained in Ref.\onlinecite{Pitarke2}, which relates the
imaginary part of the noninteracting density correlation function
$Y_{q_1,q_2}^0$ with the function $H_{q,q_1}$ of Eq. (\ref{hache}):
\begin{equation}\label{relation}
{\rm Im}Y_{q_1,q_2}^0={1\over
2}\left[H_{q_1,-q_2}+H_{-q_2,q_1}+H_{-q_3,q_2}+\left(q_2\rightarrow
q_3\right)\right].
\end{equation}
Introduction of Eqs. (\ref{single2}) and (\ref{double2}) into Eqs.
(\ref{tau}) and (\ref{stopping}) yields, after some algebra,
\begin{eqnarray}\label{tau1}
\tau^{-1}=&&4\pi\,Z_1^2\int{d^4q\over(2\pi)^4}\,v_{\bf q}
\left\{-{\rm Im}K_q+4\pi\,Z_1\int{d^4q_1\over(2\pi)^4}\,v_{{\bf q}_1}
\,v_{{\bf q}-{\bf q}_1}\,\delta\left(q_1^0-{\bf q}_1\cdot{\bf
v}\right)\right.\cr\cr
&&\left.\times\left[f_1(q,q_1)+f_2(q,q_1)+f_3^a(q,q_1)+f_3^b(q,q_1)
+f_4(q,q_1)\right]
\right\}\,\delta\left(q^0-{\bf q}\cdot{\bf v}\right)\Theta(q^0)
\end{eqnarray}
and\cite{note2}
\begin{eqnarray}\label{stopping1}
-{dE\over dx}=&&{4\pi\over v}\,Z_1^2\int{d^4q\over(2\pi)^4}\,q^0\,v_{\bf q}
\left\{-{\rm Im}K_q+4\pi\,Z_1\int{d^4q_1\over(2\pi)^4}\,v_{{\bf q}_1}
\,v_{{\bf q}-{\bf q}_1}\,\delta\left(q_1^0-{\bf q}_1\cdot{\bf
v}\right)\right.\cr\cr
&&\left.\times\left[f_1(q,q_1)+f_2(q,q_1)+f_3^a(q,q_1)+f_5(q,q_1)\right]
\right\}\,\delta\left(q^0-{\bf q}\cdot{\bf v}\right)\Theta(q^0),
\end{eqnarray}
where
\begin{eqnarray}
f_1(q,q_1)&=&{\rm Im}\tilde K_q\,{\rm Re}Y_{q,-q_1}^0\,{\rm Re}\tilde K_{q_1}\,
{\rm Re}\tilde K_{q-q_1},\\
f_2(q,q_1)&=&{\rm Re}\tilde K_q\,H_{q,q_1}\,{\rm Re}\tilde K_{q_1}\,{\rm
Re}\tilde K_{q-q_1},\\
f_3^a(q,q_1)&=&-2\,{\rm Im}\tilde K_q\,H_{q_1,q}\,{\rm Im}\tilde K_{q_1}\,{\rm
Re}\tilde K_{q-q_1},\\
f_3^b(q,q_1)&=&-{\rm Re}\tilde K_q\,H_{q,q_1}\,{\rm Im}\tilde K_{q_1}\,{\rm
Im}\tilde K_{q-q_1},\\
f_4(q,q_1)&=&-{1\over 3}\,{\rm Im}\tilde K_q\,{\rm Re}Y_{q,-q_1}^0\,{\rm
Im}\tilde K_{q_1}\,{\rm Im}\tilde K_{q-q_1}
\end{eqnarray}
and
\begin{equation}
f_5(q,q_1)={\rm
Im}\left(\tilde K_q\,\tilde
K_{q_1}^*\,\tilde K_{q-q_1}\right)\,J_{q-q_1,-q_1}.
\end{equation}
Within RPA,
the inverse dielectric functions $K_q$ and
$\tilde K_q$ coincide, and 
Eq. (\ref{stopping1}) reduces to the result of
Ref. \onlinecite{Pitarke1}. 

Eq. \ref{tau1} has not been reported
before, even within the RPA approximation.
In next section we will show that it is equivalent to
the result reported in Ref. \onlinecite{RP00}, where a derivation of
the decay rate  as the imaginary part of the on-shell self-energy of the
probe particle was sketched briefly.

\subsection{Self-energy approach}

Since we are considering the interaction of a moving probe particle with a
spatially uniform electron gas, invariant under translations, the exact
probe-particle propagator can be written in the form of an algebraic Dyson's
equation\cite{Fetter}
\begin{equation}\label{disonion}
D_{p}=D_{p}^0+D_p^0\,\Sigma_p\,D_p,
\end{equation}
which defines the self-energy $\Sigma_p$ of the probe particle. With the aid
of Eq. (\ref{d0}), Dyson's equation can be solved explicitly as
\begin{equation}
D_p={1\over p^0-\omega_{\bf p}-\Sigma_p+i\eta}.
\end{equation}

The energy and lifetime of the  excited state (quasiparticle) obtained by
adding a particle to an interacting ground state are determined by the poles
of the analytical continuation of the one-particle Green function. Hence, the
energy of the quasiparticle is $\omega_{\bf p}+{\rm Re}\Sigma_p$, and the
probability for it to occupy a given excited state decays
exponentially in time with the decay constant
\begin{equation}\label{tau0}
\tau^{-1}=-2\,{\rm Im}\Sigma_p,
\end{equation}
with the self-energy calculated at the pole of the one-particle
propagator $D_p$.

The self-energy can be represented diagrammatically as the sum of the
so-called proper self-energy insertions, i.e., all Feynman diagrams that cannot
be separated into two pieces by cutting a single particle line. Since the probe
particle, of charge $Z_1$, is assumed to be distinguishable from the electrons in
the Fermi sea, the self-energy may be expanded in powers of $Z_1$, diagrams of
order $Z_1^n$ containing $n-1$ probe-particle propagators. For a homogeneous
electron gas, contributions from the uniform positive background are canceled
by the sum of the so-called 'tadpole' diagrams; therefore, after
resuming all electron-loop corrections, the self-energy of
the probe particle can be represented diagrammatically up to third order in
$Z_1$ as in Fig. 5. The sum of the first two diagrams represents the so-called
GW approximation, and the third diagram accounts for $Z_1^3$ corrections to the
decay rate of the quasiparticle. One finds:
\begin{equation}\label{auto1}
\Sigma_p=i\,Z_1^2\int{dq^4\over(2\pi)^4}\,v_{\bf q}\,D_{p-q}
\left[(1+\chi_q\,v_{\bf q})-2\,i\,Z_1\int{d^4q_1\over(2\pi)^4}\,
D_{p-q_1}\,D_{p-q+q_1}\,Y_{q,-q_1}\,v_{{\bf q}_1}v_{{\bf q}-{\bf q}_1}\right],
\end{equation}
where $\chi_q$ and $Y_{q_1,q_2}$ represent the {\it exact} density
correlation functions of the interacting FEG, as obtained from Eqs.
(\ref{eq17}) and (\ref{eq2p}), respectively.

If the probe particle is an ion ($M>>1$), the propagator $D_p$ and
the energy $p^0$ entering Eq. (\ref{auto1}) can be safely approximated by the
noninteracting propagator $D_p^0$ and energy $\omega_{\bf p}$. Furthermore,
recoil can be neglected, and one easily finds
\begin{equation}\label{feyn}
D_{p-q}^0=-{1\over q^0-{\bf q}\cdot{\bf v}-i\,\eta}.
\end{equation}
In order to exploit the symmetry properties of $Y_{q_1,q_2}$ it is
useful to rewrite the retarded probe-particle propagator in terms of
its Feynman version as follows, 
\begin{equation}\label{new1}
D_{p-q}^0=-{1\over q^0-{\bf q}\cdot{\bf v}+i\,\eta_q}-
2\,i\,\pi\,\delta(q^0-{\bf q}\cdot{\bf v})\,\Theta(q^0). 
\end{equation}
Introducing Eq. (\ref{new1}) into Eq. (\ref{auto1}) and noting that the
time-ordered density correlation functions $\chi_q$ and $Y_{q_1,q_2}$ are
invariant under the changes ($q^0\rightarrow -q^0$) and
($q_1^0\rightarrow-q_1^0$,
$q_2^0\rightarrow -q_2^0$), respectively, some work of rearrangement leads us to
the following expression:
\begin{eqnarray}\label{auto2}
\Sigma_{{\bf p},\omega_{\bf p}}=2\,\pi\,Z_1^2\int{dq^4\over(2\pi)^4}\,
&&v_{\bf q}
\left[(1+\chi_q\,v_{\bf q})-{4\over 3}\,\pi\,Z_1\int{d^4q_1\over(2\pi)^4}\,
Y_{q,-q_1}\,v_{{\bf q}_1}v_{{\bf q}-{\bf q}_1}\,
\delta(q^0_1-{\bf q}_1\cdot{\bf v})\right]\cr\cr
&&\times\,\delta(q^0-{\bf q}\cdot{\bf v})\,\Theta(q^0).
\end{eqnarray}

Within RPA, the density correlation functions $\chi_q$ and $Y_{q_1,q_2}$ are
those given by Eqs. (\ref{RPA1}) and (\ref{RPA2}). Beyond RPA, they are
obtained from Eqs. (\ref{chitilde}) and (\ref{Ytilde}), in terms of the
noninteracting density correlation functions [$\chi_q^0$ and $Y_{q_1,q_2}^0$] and
the effective e-e interaction of Eq. (\ref{eqHub}). Hence, introduction of Eq.
(\ref{auto2}) into Eq. (\ref{tau0}) yields the following expression for the
decay rate:
\begin{eqnarray}\label{tau2}
\tau^{-1}=&&4\pi\,Z_1^2\int{dq^4\over(2\pi)^4}\,v_{\bf q}\,\delta(q^0-{\bf
q}\cdot{\bf v})\,\Theta(q^0)\cr\cr
&&\times\left[-{\rm Im}K_q+{4\over 3}\,\pi\,Z_1\int{d^4q_1\over(2\pi)^4}\,
{\rm Im}\left(\tilde K_q\,Y_{q,-q_1}^0\,\tilde K_{q_1}\,\tilde
K_{q-q_1}\right)\,v_{{\bf q}_1}v_{{\bf q}-{\bf q}_1}\,
\delta(q_1^0-{\bf q}_1\cdot{\bf v})\right],
\end{eqnarray}
where $K_q$ and $\tilde K_q$ represent the inverse dielectric functions of Eqs.
(\ref{inverse}) and (\ref{inversep}), with the density
correlation function $\chi_q$ being given in both cases by Eq.
(\ref{chitilde}). 

The equivalence of Eqs. (\ref{tau1}) and (\ref{tau2}) follows from the
expansion of the imaginary part of $\tilde K_q\,Y_{q,-q_1}^0\,\tilde
K_{q_1}\tilde K_{q-q_1}$ in Eq. (\ref{tau2}), and the use of Eq.
(\ref{relation}) and the symmetry properties of $Y^0_{q_1,q_2}$ and $\tilde
K(q^0, {\bf q})$. However, while (\ref{tau1}) has been derived by only
introducing self-energy and vertex insertions that can be described with the use
of a static local-field factor, we have now demonstrated that either Eq.
(\ref{tau1}) or Eq. (\ref{tau2}) can be used with  inclusion of many-body
dynamic local-field corrections. 

Finally, we note that although both Eq. (\ref{tau1}) [derived from Eq.
(\ref{tau})] and Eq. (\ref{tau2}) [derived from Eq. (\ref{tau0})] represent the
total decay rate, the integrands of these integral representations do not
necessarily coincide with the probability
$\left[\gamma_q^{single}+\gamma_q^{double}+...\right]$ for the probe particle to
transfer four-momentum $q$ to the FEG. Consequently, the stopping power of the
FEG for the probe particle [see Eq. (\ref{stopping1})] cannot be obtained by
simply inserting $q^0/v$ inside the integral of Eq. (\ref{tau1}) or Eq.
(\ref{tau2}), and the knowledge of the self-energy alone is not, therefore,
sufficient to calculate the stopping power. 

\subsection{Quadratic response}

In Ref. \onlinecite{RP00} the stopping power of a heavy probe-particle
was calculated using the framework of quadratic response theory. It
was found that
\begin{eqnarray}\label{eloss2}
-{dE\over dx}=&&4\pi\,Z_1^2\int{d^4 q\over(2\pi)^4}\,q^0\,v_{\bf
q}\,\delta\left(q^0-{\bf q}\cdot{\bf v}\right)\,\Theta(q^0)\cr\cr &&\times
\left[-{\rm Im}K_q^R+2\pi\,Z_1\int{d^4 q_1\over(2\pi)^4}\,{\rm
Im}\left(\tilde
K_q^R\,Y_{q,-q_1}^{R,0}\,\tilde K_{q_1}^R\,\tilde K_{q-q_1}^R\right)\,v_{{\bf
q}_1}\,v_{{\bf q}-{\bf q}_1}\,\delta\left(q_1^0-{\bf q}_1\cdot{\bf
v}\right)\right],
\end{eqnarray}
where $K_q^R$, $\tilde K_q^R$ and $Y^R_{q_1,q_2}$ represent the retarded counterparts
of $K_q$, $\tilde K_q$ and $Y_{q_1,q_2}$.

In order to compare this result to the one quoted in Sec.2 we must recall the
relationship between the time-ordered and retarded functions. 

In our case of interest, differences in the inverse dielectric functions arises from
differences in the FEG density correlation function $\chi^0$. We have:
\begin{equation}
{\rm Re}\,\chi_q^{R,0}={\rm Re}\,\chi_q^0,
\end{equation}
\begin{equation}
{\rm Im}\,\chi_q^{R,0}={\rm sgn}(q^0)\,{\rm Im}\,\chi_q^0,
\end{equation}

The relationship between $Y_{q_1,q_2}$ and $Y^R_{q_1,q_2}$ is best
analyzed using their spectral representations. $Y^R_{q_1,q_2}$ has
the same structure as Eq. (\ref{eq2p}), with $\eta_q$
replaced by a positive $\eta$. \cite{Esbensen} 
This property leads to  the following relations between
imaginary and real part of the time-ordered and
retarded $Y^0_{q1,q2}$ functions:
\begin{equation}\label{relationr}
{\rm Re}\left(Y_{q_1,q_2}^0-Y_{q_1,q_2}^{R,0}\right)=J_{-q_2,q_3}+J_{-q_3,q_2},
\end{equation}
and
\begin{equation}\label{relationim}
{\rm Im}\,Y_{q_1,q_2}^{R,0}={1\over
2}\left[{\rm
sgn}(q_1^0)\,H_{q_1,-q_2}-{\rm
sgn}(q_2^0)\,H_{-q_2,q_1}-{\rm
sgn}(q_3^0)\,H_{-q_3,q_2}+\left(q_2\rightarrow q_3\right)\right].
\end{equation}

After some work of rearrangement, and taking into account the symmetry
properties of the functions involved, we find that Eq. (\ref{eloss2}) coincides exactly with Eq.
(\ref{stopping1}). As in the case of the decay rate of Eq.
(\ref{tau2}), we find  that both Eqs. (\ref{stopping1}) and
(\ref{eloss2}) can be used with  inclusion of many-body dynamic local-field
corrections. We also note that within RPA both Eqs. (\ref{stopping1}) and
(\ref{eloss2}) reduce to the result derived in
Refs.\onlinecite{Sung,Zaremba,Esbensen,Pitarke1,Pitarke2}.\cite{note3}

\section{Conclusions}

We have developed various many-body theoretical approaches to the quadratic
decay rate and energy loss  of charged particles moving in an
electrons gas, with inclusion of short-range XC effects.

We have carried out a perturbative formulation of the scattering
matrix to derive general expressions for both the total decay rate and the
average energy loss of arbitrary moving charged particles that are
distinguishable from the electrons in the Fermi gas. Simplified expressions for
these quantities have been obtained in the case of heavy probe particles
($M>>1$). The total decay rate of heavy particles
has then been rederived from the knowledge of the probe-particle self-energy
and it has been proved that the stopping power of the heavy particle 
agrees with the result
deduced using quadratic response theory. 
Comparison of the different formalisms for a heavy particle suggests that
our results in the scattering 
formalism can be used
with full inclusion of many-body dynamic local-field corrections. 

It has also
been shown that while the first-order contributions to the energy loss may be
obtained from the total decay rate by simply inserting the energy transfer inside
the integrand of this quantity, this procedure cannot be generalized to the
description of the second-order energy loss. Since response theory is
only valid for heavy particles, this implies that the
stopping power of light particles must be calculated using scattering
theory.

\acknowledgments

The authors acknowledge partial support by the University of the Basque
Country, the Basque Hezkuntza, Unibertsitate eta Ikerketa Saila, and the
Spanish Ministerio de Educaci\'on y Cultura.

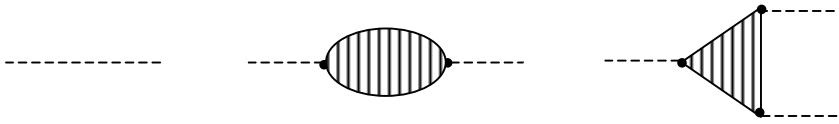
\begin{figure}
\caption{(a) Direct, (b) linear, and (c) quadratic contributions to the
screened interaction. Dashed lines represent the bare Coulomb interaction,
$-i\,v_{\bf q}$. Two- and three-point loops represent time-ordered
density correlation functions, $i\,\chi_q$ and $-2\,Y_{q_1,q_2}$, respectively.}
\end{figure}

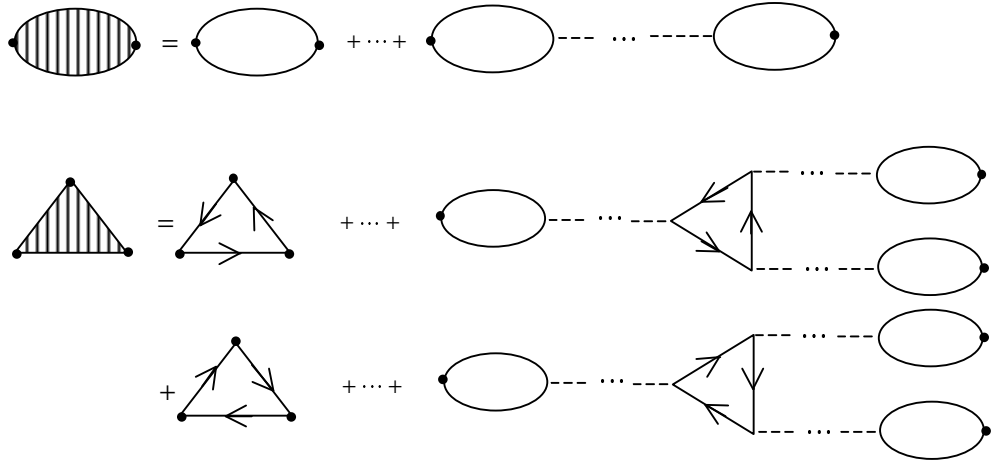
\begin{figure}
\caption{(a) The interacting RPA two-point density correlation
function, represented by a full bubble, is obtained by summing over the
infinite set of diagrams that contain a string of empty bubbles. (b) The
interacting RPA three-point density correlation function, represented by a
full triangle, is obtained by summing over the infinite set of diagrams that
combine three strings of empty bubbles through an empty triangle.}
\end{figure}

\begin{figure}
\caption{(a) Self-energy, (b) vertex, and (c) ladder insertions, which are
neglected within RPA.}
\end{figure}

\begin{figure}
\caption{Diagrammatic representation of the scattering-matrix elements of (a)
Eq. (\ref{tsinglep}) and (b) Eq. (\ref{tdoublep}). Thick and thin solid lines
represent noninteracting probe-particle and electron propagators, $i\,D_p^0$ and
$i\,G_k^0$, respectively. Wiggly lines represent the screened e-e Coulomb
interaction, 
$-i\,\tilde v_{\bf
q}\,\tilde K_q$, and thick discontinuous lines represent the screened
interaction between electrons and probe particle, $-i\, v_{\bf
q}\,\tilde K_q$}
\end{figure}

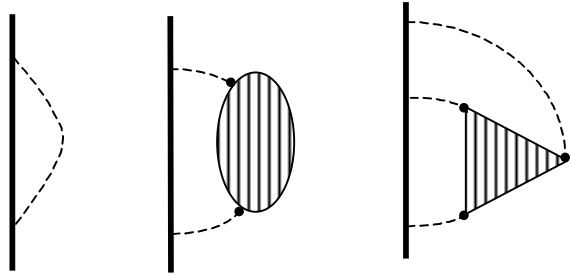
\begin{figure}
\caption{The probe-particle self-energy, up to third order in $Z_1$. Thick
solid lines represent the exact probe-particle propagator, $i\,D_p$. Dashed
lines represent the bare Coulomb interaction, $-i\,v_{\bf q}$. Two- and three-point loops represent time-ordered
density correlation functions, $i\,\chi_q$ and $-2\,Y_{q_1,q_2}$, respectively.}
\end{figure}


\begin{references}
\bibitem{note1} Contributions to the probability $\gamma_q^{\rm double}$ that
are proportional to $Z_1^2$ are not included here, since they all represent
short-range correlations not present in the RPA.
\bibitem{note2} The term $f_5(q,q_1)$ of Eq. (\ref{stopping1}) is missing in
Eq. (4.13) of Ref.\onlinecite{Pitarke2}. This contribution to the energy loss
is found to be negligible at low and high velocities, and the inclusion of this
term results, within RPA, in a $Z_1^3$ correction to the stopping power that is
at intermediate velocities lower than that reported in Ref.\onlinecite{Pitarke2}
by less than $10\%$. We note that the symmmetrized empty three-point function
defined in Ref.\onlinecite{Pitarke2} is $M_{q,q_1}=-Y_{q,-q_1}^0$.
\bibitem{note3} In Ref.\onlinecite{Pitarke2}, the difference between the real
parts of $Y_{q_1,q_2}^0$ and $Y_{q_1,q_2}^{R,0}$ [see Eq. (\ref{relationr})] was
overlooked. As a result, the energy loss of Eq. (\ref{eloss2}) was found, within
RPA, to be equivalent to that of Eq. (\ref{stopping1}), but with no inclusion
of the term $f_5(q,q_1)$. This term is now found when Eq. (\ref{relationr}) is
taken into account.
\end{references}
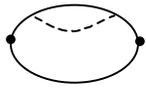
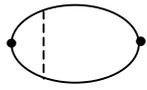
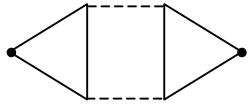
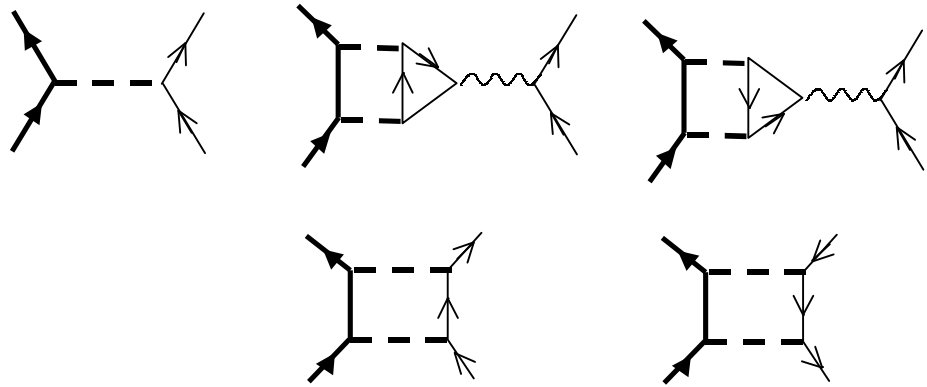
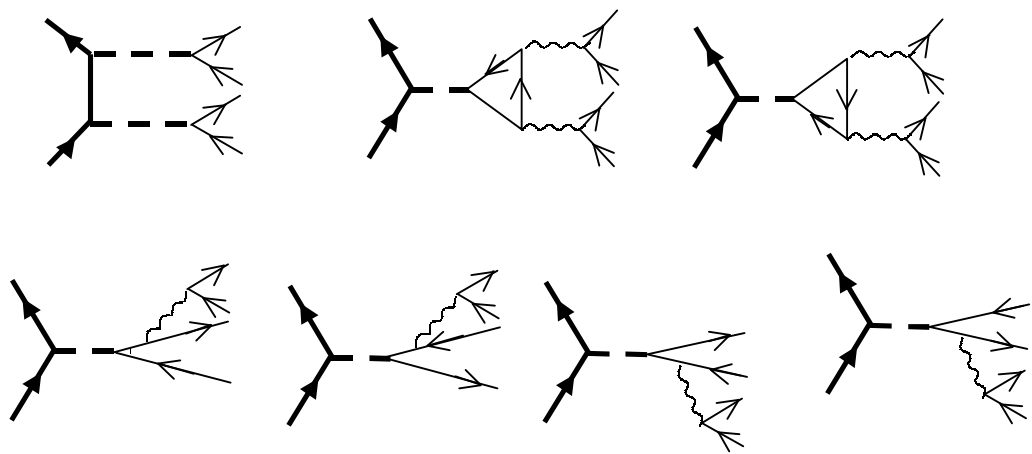
\end{document}